\documentclass[aps,prl,twocolumn,superscriptaddress,floatfix]{revtex4-2}
\usepackage{blindtext}
\usepackage{amsmath}

\usepackage{hyperref}
\hypersetup{
  colorlinks   = true, 
  urlcolor     = blue, 
  linkcolor    = blue, 
  citecolor   = blue 
}

\usepackage{bm}
\usepackage{gensymb}
\usepackage[normalem]{ulem}
\usepackage{euscript}
\usepackage[pdftex]{graphicx}
\usepackage{xspace}
\usepackage{xfrac}
\usepackage{enumerate}
\usepackage{braket}
\usepackage{float,fancyvrb}
\usepackage[T1]{fontenc}
\usepackage{lmodern}
\graphicspath{{./Figures/}}
\usepackage{wrapfig}
\usepackage{scrextend}
\usepackage{comment}
\usepackage{url}
\usepackage{amssymb}
\usepackage{xcolor}

\def\be{\begin{equation}}
\def\ee{\end{equation}}
\def\beq{\begin{eqnarray}}
\def\eeq{\end{eqnarray}}
\def\ba#1{\begin{array}{#1}}
\def\ea{\end{array}}
\def\bn{\begin{enumerate}}
\def\en{\end{enumerate}}

\definecolor{ao}{rgb}{0.0, 0.5, 0.0}

\begin{document}

\title{Markovian baths and quantum avalanches}

\author{Dries Sels}
\affiliation{Department of Physics, New York University, New York, NY, USA}
\affiliation{Center for Computational Quantum Physics, Flatiron Institute, New York, NY, USA}
\date{\today}

\begin{abstract}
In this work I will discuss some numerical results on the stability of the many-body localized phase to thermal inclusions. The work simplifies a recent proposal by Morningstar et al. [arXiv:2107.05642] and studies small disordered spin chains which are \emph{perturbatively} coupled to a Markovian bath. The critical disorder for avalanche stability of the canonical disordered Heisenberg chain is shown to exceed $W^\ast \gtrsim 20$. In stark contrast to the Anderson insulator, the avalanche threshold drifts considerably with system size, with no evidence of saturation in the studied regime. I will argue that the results are most easily explained by the absence of a many-body localized phase. 
\end{abstract}

\maketitle

To date, the putative many-body localized (MBL) phase is the sole example of a generic interacting local Hamiltonian system that fails to thermalize under its own internal dynamics. In contrast to other non-ergodic models, the integrability of the MBL phase is \emph{emergent}. As such, it has attracted a lot attention from the community over the past decade~\cite{nandkishore2015many, abanin2019review}. After the initial proposal, research has been primarily focused on understanding the nature of the phase transition between the thermal and the MBL phase.

Recently, a number of works have questioned the stability of the MBL phase~\cite{suntajs2019quantum,suntajs2020transition, Kiefer_Emmanouilidis_2021,sels2020dynamical,sels2021thermalization,vidmar2021phenomenology}. In turn, some of the conclusions were challenged by follow up papers~\cite{abanin2019rebuke, Sierant_2020,LuitzReply,Panda_2020,crowley2021constructive}. Although some of the critique in the aforementioned works is technical, in essence they question the validity of the conclusions drawn in ref.~\cite{suntajs2019quantum,suntajs2020transition, Kiefer_Emmanouilidis_2021,sels2020dynamical,sels2021thermalization} because most of the numerics is done in, or in the vicinity of, the thermal phase.

In this context, it would be highly desirable to numerically investigate the instability of the MBL phase, rather than its emergence from the thermal phase. Direct numerical verification of incipient ergodicity is very challenging. Even for clean systems, in which there is a tradition of studying the approach to eigenstate thermalization, probing the onset of chaos has been notoriously hard. The leading proposed instability of MBL is based on so called \emph{avalanches} induced by rare thermal inclusions. Exact numerical studies are limited to small system sizes and can therefore not directly capture these inclusions. To circumvent this bottleneck I adopt a recent proposal by Morningstar et al.~\cite{morningstar2021avalanches}. Instead of directly investigating rare thermal regions in a closed systems, let us study the transient dynamics of a system coupled to an infinite bath. For an infinite bath the system will definitely thermalize, but within the avalanche theory~\citep{DeRoeck2017avalanche,thiery2017microscopically} it is the rate of thermalization that governs how large the thermal inclusions become. The argument is rather simple, consider a thermal inclusion of size $\ell_0$, and imagine it has been able to thermalize $\ell$ spins on both sides, then the inclusion has grown to a size $\ell_0+2\ell$. Consequently, the level spacing of the inclusion has now become $\sim 2^{-(\ell_0+2\ell)}$. The inclusion can only serve as a proper bath as long as it thermalizes spins at a rate $\Gamma$ which exceeds the level spacing, such that the spins can not resolve the discreteness of the spectrum.  For a finite size system of length $L$, the critical thermalization rate to be stable against avalanches thus becomes $\Gamma \lesssim 4^{-L}$. The main purpose of this paper is to investigate the behavior of this thermalization rate $\Gamma$. Before going on to describe the exact procedure, it's interesting to note that, coming from the ergodic side, ergodicity gets broken when the Thouless time exceeds the Heisenberg time, i.e. when $\Gamma \lesssim 2^{-L}$. This opens up an entire parameter regime of disorder strengths where a finite size systems lacks complete ergodicty but it is still unstable to avalanches. I would argue that large parts of the MBL literature, in particular works with numerical support, have been conducted in this intermediate regime. 

\emph{Method--}
To study the relaxation rate, consider a system coupled to infinite Markovian bath such that the dynamics can be described by the following master equation 
\begin{equation}
\partial_t \rho= \mathcal{L}(\rho),
\label{eq:Markov}
\end{equation}
with a Liouvillian super-operator
\begin{equation}
\mathcal{L}(\rho)=-i[H,\rho]+ \gamma \sum_\mu \left(L_\mu \rho L^\dagger_\mu -\frac{1}{2} \left\lbrace L_\mu L^\dagger_\mu,\rho \right\rbrace  \right).
\label{eq:Liouv}
\end{equation}
Here, $H$ denotes the bare Hamiltonian of the system, $L_\mu$ specifies the nature of the coupling to the bath and $\gamma$ simply denotes the nominal strength of the coupling. In what follows, I will consider the canonical disordered Heisenberg model 
\begin{equation}
H= \frac{1}{4 }\sum_{i=1}^{L-1} \vec{\sigma}_i\cdot \vec{\sigma}_{i+1}+ \frac{1}{2} \sum_{i=1}^L h_i Z_i,
\end{equation}
with $h_i$ being i.i.d. random variables drawn out of uniform distribution on $[-W,W]$ and $\vec{\sigma}_i=(X_i,Y_i,Z_i)$ the vector composed of Pauli operator. To mimic the thermal inclusion, the Lindblad operators are taken to be the Pauli operators on the first spin, i.e. $L_\mu=(X_1,Y_1,Z_1)$. The solution to the Markovian master equation~\eqref{eq:Markov} is formally given by
\begin{equation}
\rho_t=\sum_i  e^{\lambda_i t} p_i \rho_i,
\end{equation}
where $\lambda_i$ are the eigenvalues and $\rho_i$ the right eigenoperators of the Liouvillian. The coefficients $p_i$ denote the overlap of the initial state with the left eigenoperators of the Liouvillian. The stationary state, $\lambda_0=0$, will be unique and given by the infinite temperature ensemble $\rho_0 \propto \mathbb{I}$. The slowest decaying operator is associated with the eigenvalue with the second largest real part $\lambda_1$, its  relaxation rate is given by $\Gamma=-{\rm Re}(\lambda_1)$. 

Finding the exact spectrum of the Liouvillian super-operator is numerically quite demanding. It requires diagonalizing a $4^N \times 4^N$ matrix for a system composed of $N$ spins and is thus limited to very small systems. This is where I deviate from Morningstar et al.~\cite{morningstar2021avalanches}. In ref.~\cite{morningstar2021avalanches} the coupling to the bath, $\gamma$ in eq.~\eqref{eq:Liouv}, is taken to be of $O(1)$. However, at present, there is no immediate reason to be interested in non-perturbative effects of the system-bath coupling. In fact, at sufficiently large system-bath coupling one will induce a Zeno effect, resulting in a relaxation rate $\Gamma \propto 1/\gamma$. The virtue of working at weak coupling is that the decay rate can be computed perturbatively. For $\gamma=0$, the eigenoperators and eigenvalues of the Liouvillian are simply given $\rho_{nm}= \left|n\right> \left< m \right|$ and $\lambda_{nm}=i(E_m-E_n)$ respectively, where $\left|n\right>$ and $E_n$ denote the eigenvectors and eigenvalues of the bare Hamiltonian of the system. Assuming the Hamiltonian has no degenerate gaps, all the $\lambda_{nm}$ are unique as long as $n\neq m$. There is an exponentially degenerate sector composed of all operators that are diagonal in the Hamiltonian, i.e.  $n=m$ such that $\lambda_{nn}=0$. 
It's easy to show that the slowest operator will be in this degenerate diagonal sector. For example, the perturbative decay rate of $\rho_{nm}$ becomes $\Gamma_{nm}= \gamma \sum_\mu  R_\mu$, with 
\begin{equation}
R_\mu =\frac{1}{2}\left(\left<n|L^2_\mu| n \right>+ \left<m|L_\mu^2 |m \right>
- 2\left<n|L_\mu|n \right> \left<m|L_\mu|m \right> \right) \nonumber.
\end{equation}
It follows that $\Gamma_{nm} \geq (\Gamma_{mm} +\Gamma_{nn})/2$, implying one of the two diagonal states must have a smaller decay rate than the off-diagonal combination. The slowest operator thus resides in the diagonal sector and can be found by diagonalizing the matrix 
\begin{equation}
M_{n,m}=\gamma \sum_\mu \left(\left<n|L^2_\mu| n \right> \delta_{n,m}-|\left<n|L_\mu|m \right>|^2 \right),
\end{equation}
such that the decay rate is given by the smallest non-zero eigenvalue of $M$. Deep in the MBL phase, the latter can be thought of as the lifetime of the most decoupled l-bit. The computational cost has been reduced to diagonalizing a $2^N\times 2^N$ (non-sparse) matrix. In what follows, I will discuss results for system sizes ranging from $L=4$ to $L=14$. Currently, numerics is limited by the precision with which one can represent the matrix elements of $M$. 

\begin{figure}[htb]
	\centering
	\includegraphics[width= 0.45\textwidth]{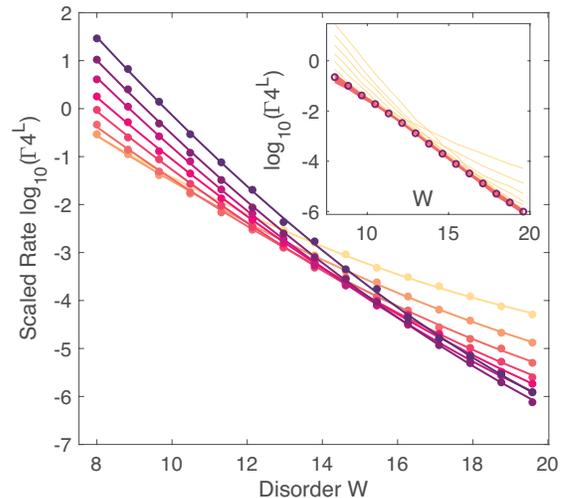}
	\caption{\textbf{Scaled decay rate:} Curves show the 80th percentile of the distribution of the smallest (non-zero) eigenvalue of the Liouvillian super-operator in an XXX chain over disorder realizations. By rescaling with $4^L$ the crossing point in the data becomes indicative of the avalanche stability threshold. Different curves show different system sizes, ranging from $L=7-14$. The crossing point drifts substantially from around $W^\ast \approx 8$ for the smallest system to $W^\ast>20$ for the largest available system size.  Inset shows the numerically extracted minimal value of $\Gamma 4^L$ with $99\%$ bootstrap confidence intervals. }
\label{fig:rate4L}
\end{figure}

\begin{figure}[htb]
	\centering
	\includegraphics[width= 0.45\textwidth]{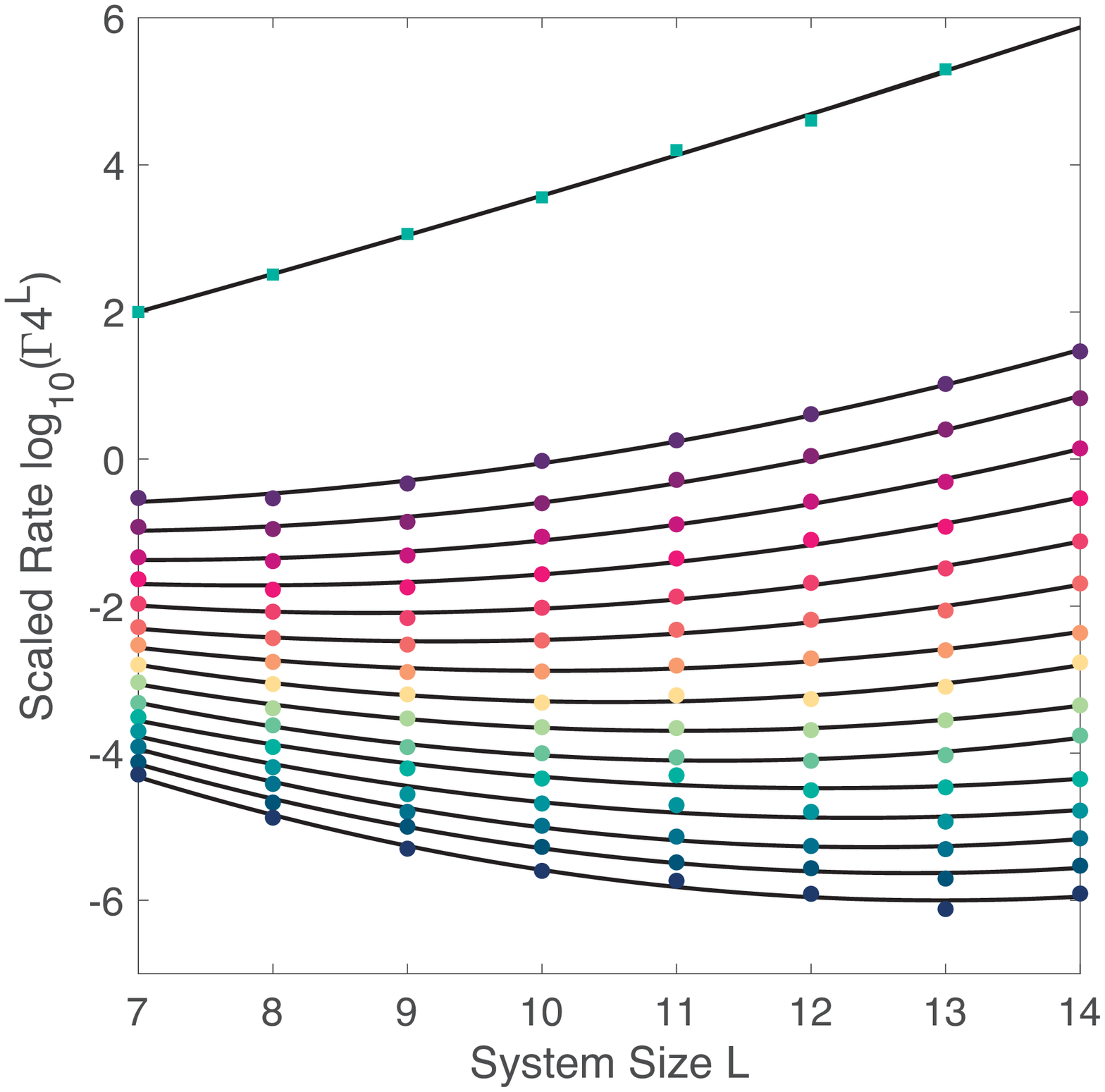}
     \includegraphics[width= 0.45\textwidth]{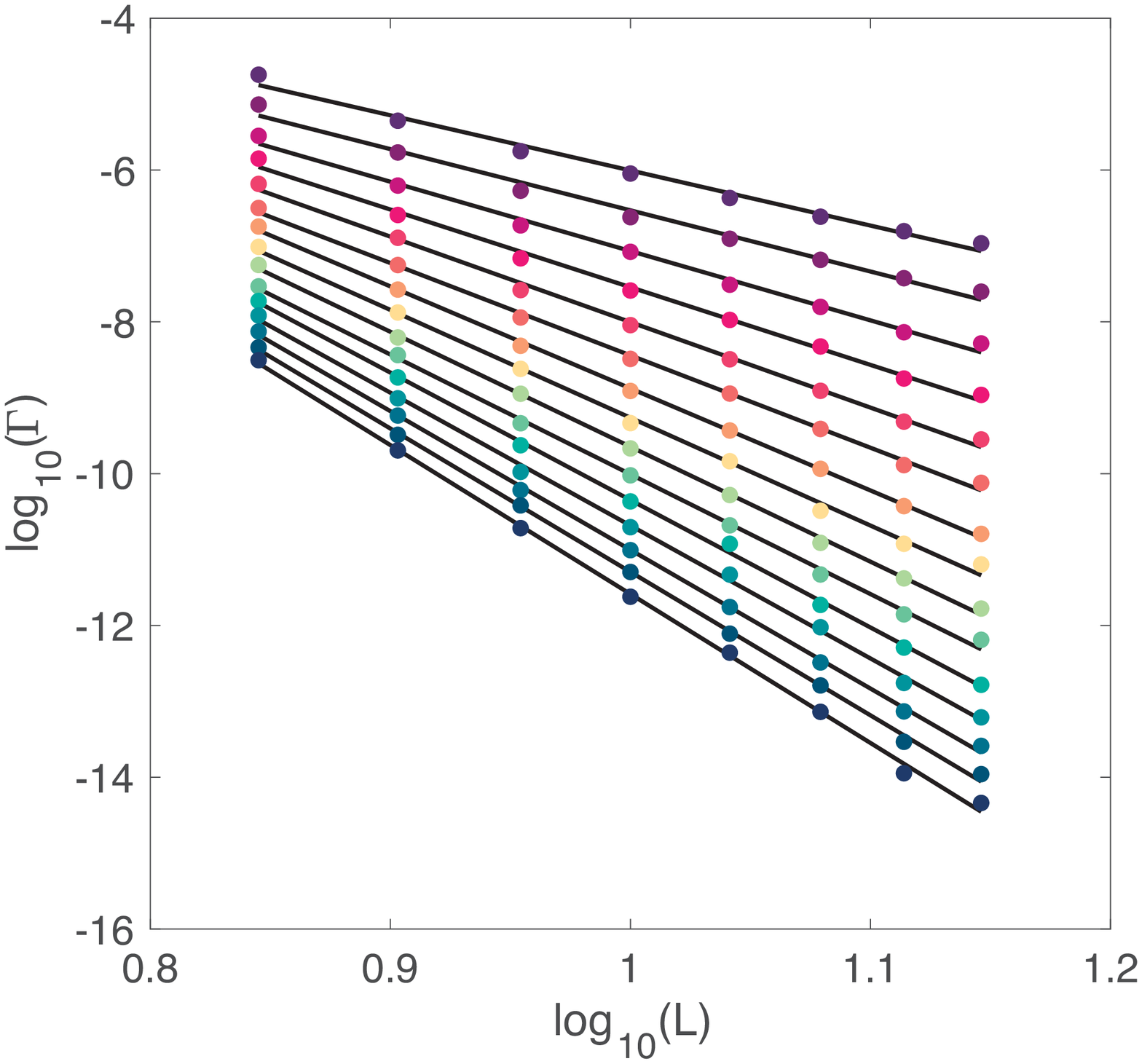}
	\caption{\textbf{Scaled decay rate II:} Curves show the behavior 80th percentile of the distribution of the smallest (non-zero) eigenvalue of the Liouvillian super-operator in an XXX chain over disorder realizations. \textbf{Panel A} Different curves correspond to different value of the disorder strength $W=8-20$; blue squares serve as a reference and correspond to weak disorder $W=1$. Black lines show quadratic fits, from which we extract the minimal value of the rescaled rate shown in the inset in Fig.~\ref{fig:rate4L}. \textbf{Panel B} shows the bare rate $\Gamma$ on a log-log scale. Black lines are linear fits, corresponding to algebraic decay of the rate with system size. The exponent $z$ is shown in Fig.~\ref{fig:exponen}. }
\label{fig:rateL}
\end{figure}

\emph{Results--} Recall that the goal is to investigate the scaling behavior of the thermalization rate $\Gamma$, in order to understand the instability of the system against avalanches. The latter occurs when the system thermalizes faster than $\Gamma \propto 4^{-L}$. Since the exact scaling form of the thermalization rate is unknown, it's instructive to look at the crossing points of the rescaled rate $g=\Gamma 4^L$ for various system sizes, as shown in Fig.~\ref{fig:rate4L}. As already noted in Ref.~\cite{morningstar2021avalanches}, the crossing point drifts substantially with system size. Extending the range of system sizes from $L=9$ (in Ref.~\cite{morningstar2021avalanches}) to $L=14$ does not result in any slowing down. In fact, the crossing point is pushed from somewhere around $W\approx 7$ to $W>20$.   This is in stark contrast to what happens in the non-interacting Anderson insulator where the crossing point barely moves (see Fig.~\ref{fig:ratescalingAnderson} in the Appendix). On the same system sizes I extract a critical point of $W\approx 1.4$ for the Anderson insulator, which is in good agreement with the $W=1.34$ obtained by Crowley and Chandran~\cite{crowleyAnderson2020}.
Furthermore, over the range of available system sizes, the crossing in Fig.~\ref{fig:rate4L} becomes shallower with increasing system size, giving the impression that there is a lower bound to the rescaled rate $g$. This minimal value can easily be extracted by looking at fixed disorder $W$ and analyzing the dependence of the data on the system size $L$, as shown in Fig.~\ref{fig:rateL}A. At sufficiently large disorder, one observes a crossover where small systems appear to become more stable to avalanches, i.e. $g$ decreases with $L$, while ultimately crossing back over to a regime in which $g$ increases with $L$. Over the available range, the data can well be approximated by a quadratic function of $L$, allowing to extract the dependence of the minimal $g$ on the disorder strength $W$. The latter appears to decay roughly exponentially with $W$ over a rather wide range. 

\begin{figure}[htb]
	\centering
	\includegraphics[width= 0.45\textwidth]{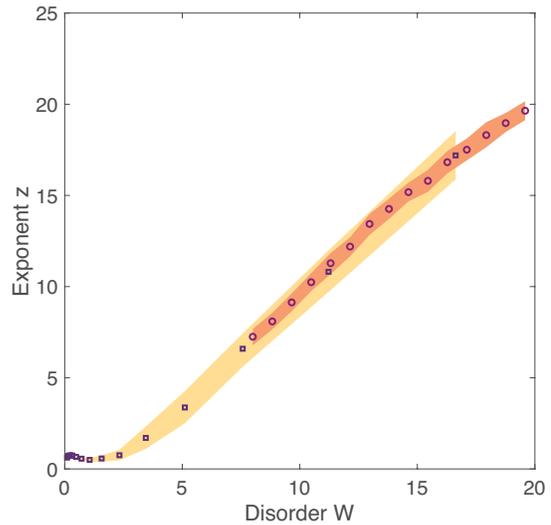}
	\caption{\textbf{Power law:} The numerically extracted exponent of the power-law decay of the thermalization rate. Shaded areas show 99$\%$ bootstrap confidence intervals. Different symbols correspond to different datasets.}
\label{fig:exponen}
\end{figure}

Needless to say that there would be no stable MBL phase in the thermodynamic limit if this behavior were to persist. The reader might rightfully be concerned about such extrapolation, and possibly even confused as to how this scaling can even hold in the regime that is being studied. After all, disorder $W>10$ was until now believed to be deep inside the MBL phase, where one would surely expect exponentially localized l-bits and as such exponentially decaying thermalization rate with $L$. On a phenomenological level, the apparent contradiction is resolved by noting that the thermalization rate does not decay exponentially but is much better described by a power-law $\Gamma \propto L^{-z(W)}$, see Fig.~\ref{fig:rate4L}B with the extracted exponent show in Fig.~\ref{fig:exponen}. Power-law scaling would of course explain the observed crossover, where systems smaller than $\sim z/\log(4)$ would appear to become more stable against avalanches. Such a powerlaw naturally complements the phenomenology from the ergodic side~\cite{sels2020dynamical,suntajs2020transition,sels2021thermalization}.

\begin{figure}[htb]
	\centering
	\includegraphics[width= 0.45\textwidth]{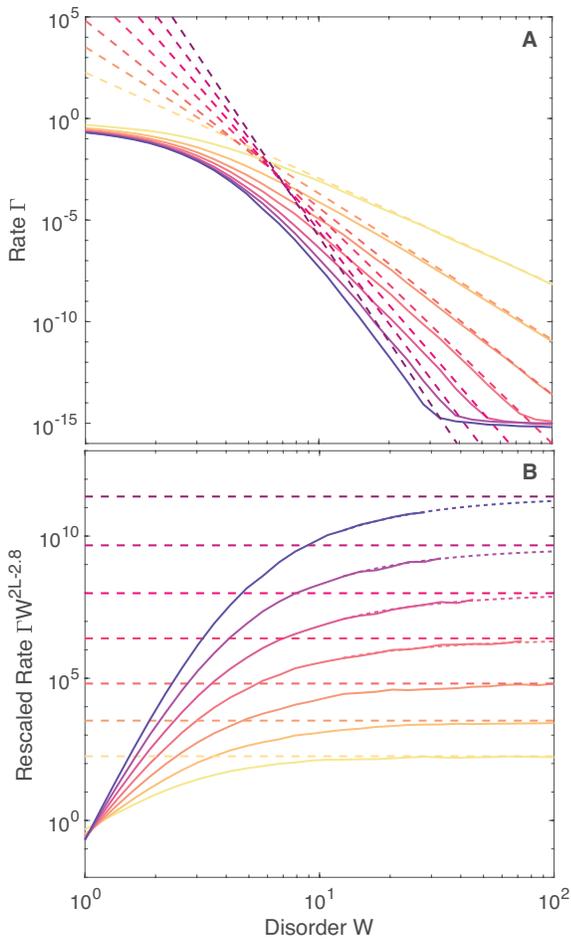}
	\caption{\textbf{Relaxation rate XXX:} The typical slowest relaxing operator in a disordered Heisenberg chain. Curves show the 80th percentile of the distribution of the smallest (non-zero) eigenvalue of the Liouvillian super-operator over disorder realizations. Different curves correspond to different systems sizes ranging from $L=4-10$. \textbf{Panel A}  shows the bare rate $\Gamma$ and \textbf{panel B} shows the rate rescaled by its expected asymptotic $W^{-2(L-1)}$ behavior. The dashed lines show the extracted asymptotes and dotted lines show the extrapolated function.}
\label{fig:ratescalingXXX}
\end{figure}

\begin{figure}[tb]
	\centering
	\includegraphics[width= 0.45\textwidth]{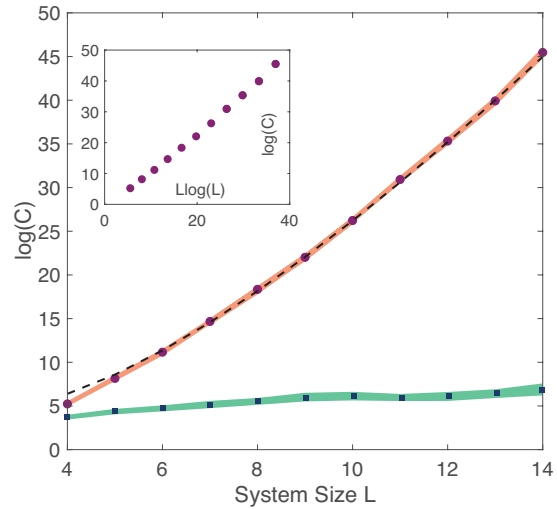}
	\caption{\textbf{Asymptotic  Scaling:} At sufficiently large disorder $W$ the slowest relaxation rate is expected to behave as $C/W^{2(L-1)}$. This figure shows the change in the numerically extracted prefactor with system size $L$. Red circles and blue squares correspond to the Heisenberg chain and free fermions respectively. Shaded areas indicate $99\%$ bootstrap confidence intervals. The black dashed line corresponds to $\propto [(L-2)!]^2$. The inset shows the same data as a function of $L\log L$. }
\label{fig:asymptC}
\end{figure}

It should be noted, however, that such scaling is not an explanation \emph{an sich}. After all, a power-law indicates some non-perturbative physics, which begs the question how this emerges out of the asymptotically large $W$ regime. Deep in the MBL phase the thermalization rate of the system should be determined by the decay of the most distant l-bit from the bath. Reaching the bath than requires using $L-1$ bonds and if all those transitions are far of resonant one would expect the thermalization rate to scale as 
\begin{equation}
\Gamma \approx \frac{C}{W^{2(L-1)}}
\label{eq:GammaAsympt}
\end{equation}
Such scaling is very different from the approximate power-law scaling and can indeed be observed at much larger $W$, see Fig.~\ref{fig:ratescalingXXX}. Due to limited numerical precision one can not access the asymptotic regime on larger system sizes. In order to extract the constant $C$, I therefore factor out the expected $W^{2L-b}$ scaling and extrapolate the data to $W \rightarrow \infty$; numerically the constant $b$ is found to be $2.8$ which seems close to the expected $b=2$. The procedure is summarized in Fig.~\ref{fig:ratescalingXXX}B. It's immediately clear that the constant $C$ in expression~\eqref{eq:GammaAsympt} has a very strong dependence on the system size. 
The extracted constant, with bootstrapped confidence intervals, is shown in Fig.~\ref{fig:asymptC}. It grows at least exponentially with system size, in contrast to the Anderson insulator in which it barely changes. In additional, there is a significant curvature on a semi-log scale, making the data much better described by a factorial rather than an exponential. This leaves two possibilities, either (i) the curvature is transient and the asymptotic behavior is simply exponential $C\sim k^{2L}$, or (ii) the factorial growth persists in the thermodynamic limit and $\log C \sim L\log L$. In the first scenario, there would be a transition at finite disorder $W^\ast \sim 2k$. 
However, even on these small system sizes I find $k>10$. While it can not be ruled out that $k$ will saturate to some larger constant, I see no reasonable explanation for such a large renormalization of the effective hopping to $J_{\rm eff}\sim k/W$; especially since it's completely absent in the Anderson insulator. 
In contrast, scenario (ii) might have a rather simple explanation. There is a key difference between the structure of perturbation theory in the interacting and non-interacting case, namely, the combinatorial structure of the latter is trivial. For example, perturbatively dressing the last spin into an l-bit, involves taking successive nested commutators of $Z$ with the Hamiltonian. The norm of $\ell$th order nested commutator of a local operator $O$ with the Hamiltonian scales as
\begin{equation}
\Vert [H,[\cdots,[H,O]]] \Vert \sim N_\ell (2J)^\ell,
\end{equation}
where $N_\ell$ keeps track of the number of connected clusters that contribute to the commutator, see ref.~\cite{Dymarsky2020}. For non-interacting systems $N_\ell=1$, as $n$-body operators remain $n$-body operators.  In one-dimensional many-body systems one can in principle asymptotically  achieve $\log(N_L)= L \log (aL/\log(bL))$ where $a$ and $b$ are constants of $O(1)$~\cite{Dymarsky2020}, highlighting the spreading through operator space. 
It should be noted that the same combinatorics is the mathematical origin of the non-zero high frequency spectral weight of local operators in many-body systems. It has recently been argued by Polkovnikov and myself that this non-zero spectral weight ultimately enables a Fermi golden rule like relaxation of spins~\cite{sels2021thermalization}. This observation thus seems to naturally merge our understanding deep on the ergodic side with that deep in the MBL regime. While this argument explains the large discrepancy between the Anderson insulator and the Heisenberg model, it's far from a proof. In order to make the statement rigorous one has to show that the energy denominators that will appear in perturbation theory do not conspire to cancel the (almost) factorial growth. 
\acknowledgements 
\emph{Acknowledgments--} I acknowledge useful discussions with D. Huse and A. Polkovnikov. The Flatiron Institute is a division of the Simons Foundation. The work is partially supported by AFOSR: Grant FA9550-21-1-0236.

\bibliography{ref_general} 

\section*{Appendix: Anderson Insulator}
This appendix studies the Anderson insulator over the same range of system sizes and disorder strengths as the results presented on the Heisenberg model in the main text. It serves as a point of comparison. 
\begin{figure}[tb]
	\centering
	\includegraphics[width= 0.45\textwidth]{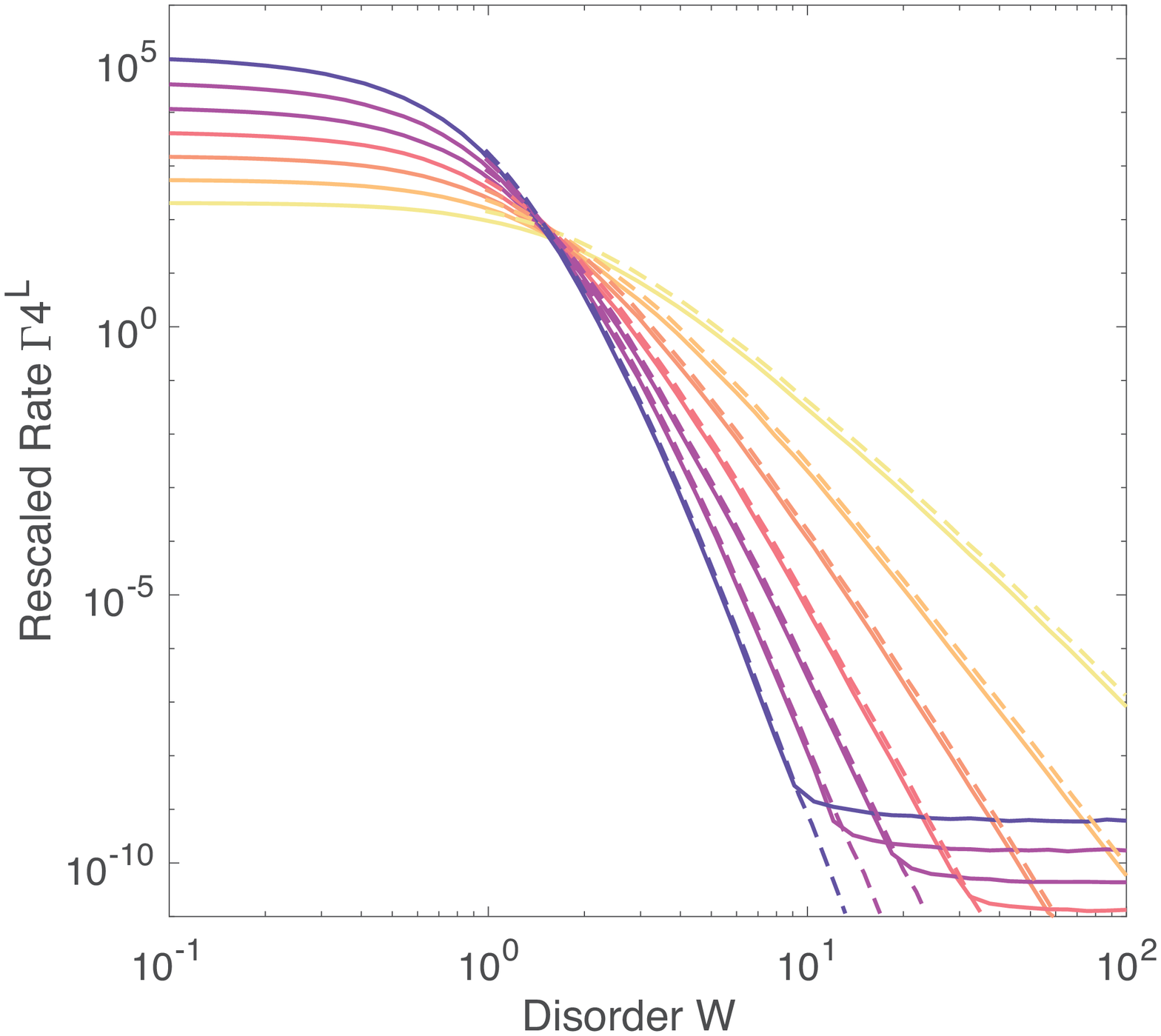}
		\includegraphics[width= 0.45\textwidth]{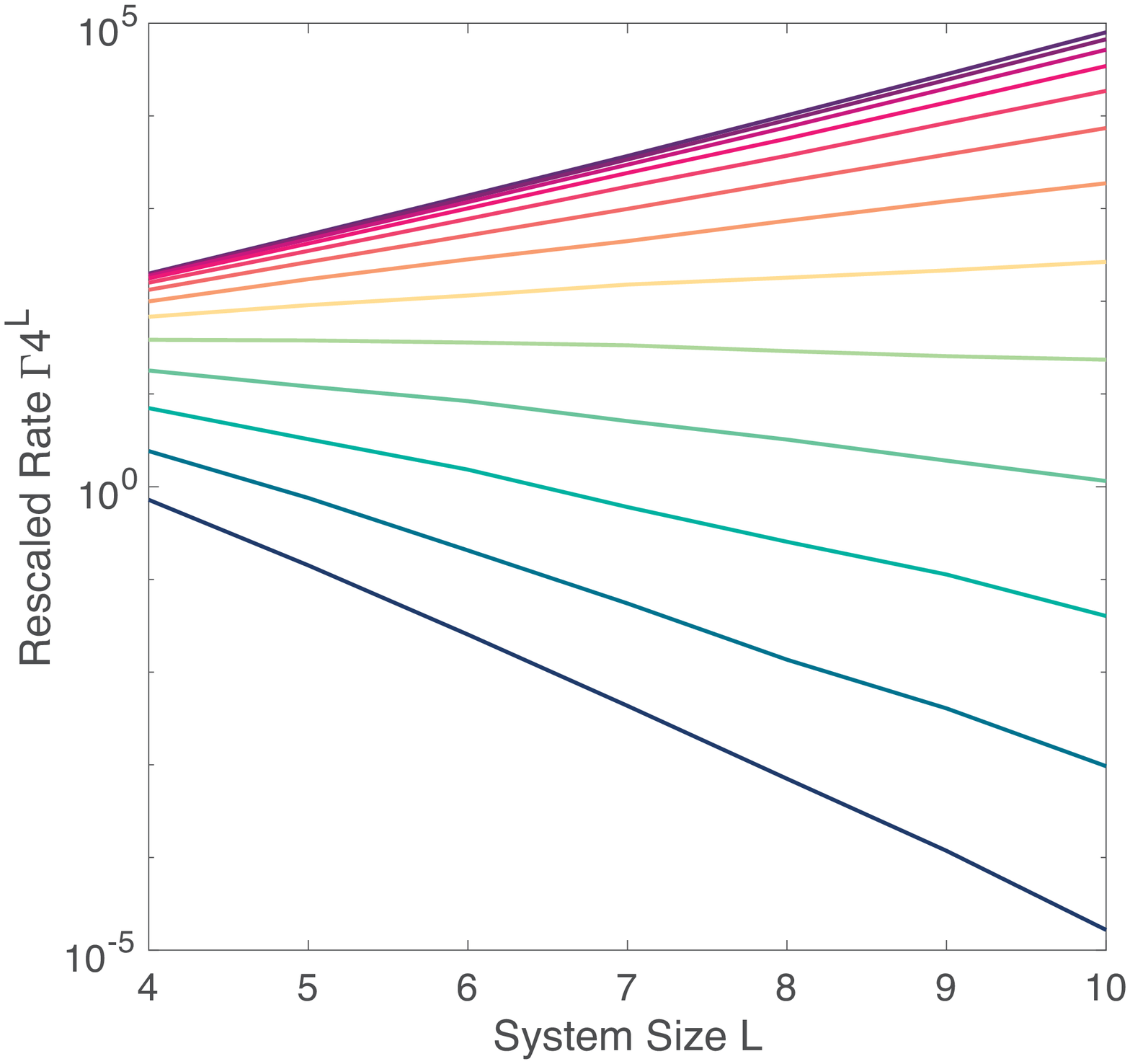}
	\caption{\textbf{Relaxation time Anderson:} The typical slowest relaxing operator in a disordered chain of non-interacting fermions. Curves show the 80th percentile of the distribution of the smallest (non-zero) eigenvalue of the Liouvillian super-operator over disorder realizations. \textbf{Panel A} Different curves correspond to different systems sizes ranging from $L=4-10$. By rescaling with $4^L$ the crossing point in the data becomes indicative of the avalanche stability threshold. A stable crossing is observed around $W \approx 1.4$. The dashed lines show the decay rate of the slowest single particle operator. \textbf{Panel B} The same data is shown as a function of $L$ for different values of the disorder strength $W$. }
\label{fig:ratescalingAnderson}
\end{figure}
Specifically, consider the Hamiltonian 
\begin{equation}
H= \frac{1}{4 }\sum_{i=1}^{L-1} (X_{i}X_{i+1}+Y_{i}Y_{i+1})+ \frac{1}{2} \sum_{i=1}^L h_i Z_i,
\end{equation}
with $h_i$ being i.i.d. random variables drawn out of uniform distribution on $[-W,W]$ and $(X_i,Y_i,Z_i)$ denoting the respective Pauli operators on each site $i$. After Jordan-Wigner transformation this maps to free fermions with on-site disorder. A detailed analysis by Crowley and Chandran~\cite{crowleyAnderson2020} concluded that the Anderson insulator should be stable against avalanches for disorder strengths $W>1.34$. Even on very small systems I find $W^\ast \approx 1.4$, with a crossing that drifts very weakly to smaller values of disorder, see Fig.~\ref{fig:ratescalingAnderson}. Note that at fixed disorder one observes a rather clean exponential growth/decay for the rescaled rate $\Gamma 4^L$, in contrast to the observed crossover behavior in the Heisenberg model.

\end{document}